\documentclass[epjCONF]{svjour}
\usepackage{graphics}
\usepackage[varg]{txfonts} 
\usepackage[latin1]{inputenc}
\session-title{A Universe of Dwarf Galaxies}
\begin{document}
\title{Thin discs, thick dwarfs, and the importance of feedback effects}
\author{R. S\'anchez-Janssen\inst{1}\fnmsep\thanks{\email{rsanchez@eso.org}} \and J. M\'endez-Abreu \inst{2,3} \and  J.A.L. Aguerri \inst{2,3}}
\institute{European Southern Observatory, Alonso de C\'ordova 3107, Vitacura, Santiago, Chile \and Instituto de Astrof\' isica de Canarias, Calle V\' ia L\'actea s/n, E-38200 La Laguna, Tenerife, Spain \and Departamento de Astrof\' isica, Universidad de La Laguna, E-38205 La Laguna, Tenerife, Spain}
\abstract{
We investigate the role of stellar mass in shaping the intrinsic thickness of faint systems by determining the probability distribution
of apparent axis ratios for two different samples that probe the faint end of the galaxy luminosity function ($M_{B} < -8$). We find that the $b/a$ distribution is a strong function of M$_{*}$, and identify a limiting stellar mass M$_{*} \approx 2\times10^{9}$ M$_{\odot}$ below which galaxies start to be systematically thicker. We argue that this is the result of the complex interplay between galaxy mass, specific angular momentum and stellar feedback effects: the increasing importance of turbulent motions in lower mass galaxies leads to the formation of thicker systems. We find a good agreement between our results and the latest numerical simulations of dwarf galaxy formation, and discuss several further implications of this finding --including the formation of bars and spirals in faint galaxies, the deprojection of H{\sc i} line profiles and simulations of environmental effects on dwarf galaxies.} 
\maketitle
\section{Introduction}
\label{intro}
Galaxy discs are key structural components in the picture of galaxy formation. Despite their enormous importance, understanding the details of disc formation remains extremely challenging, with successful models requiring the inclusion of numerous baryonic processes such as gas cooling, and heating by a cosmic UV field, star formation and supernovae \cite{G10}.
This approach, in turn, raises the question of which role does mass play in shaping the properties of discs, as these strong feedback processes are expected to produce a greater influence in lower-mass galaxies \cite{K07}.

In \cite{SJ10} we addressed this issue by investigating the range of masses where thin discs exist. Given that the intrinsic thickness of a disc population can be identified as the minimum value of the distribution of apparent axis ratios, we studied the probability distribution of $b/a$ as a function of galaxy mass (luminosity) for two different samples that probe the faint end of the galaxy luminosity function.

\section{The probability distribution of $b/a$}
\label{sec:1}
The first sample consists of \emph{all} 9245 galaxies in the SDSS-DR7 with recession velocities in the $2000 < cz < 6000$ km\,s$^{-1}$ range, thus being volume-limited for objects with $M_{i} \leq -16.5$ and including systems as faint as $M_{i} = -14.5$. 
In order to extend our analysis to even fainter luminosities, we also analysed the catalogue of neighbouring galaxies of \cite{K04}. This is an all-sky catalogue including 445 galaxies brighter than $M_{B} = -8$ with individual distance estimates $D \leq 10$ Mpc.
In both cases we have used 25 mag\,arcsec$^{-2}$ isophotal axis ratios,  which provide robust measurements of galaxy shapes at their outer regions --thus minimising the effects of dust and bulges. 

\begin{figure*}
\resizebox{0.33\textwidth}{!}{\includegraphics{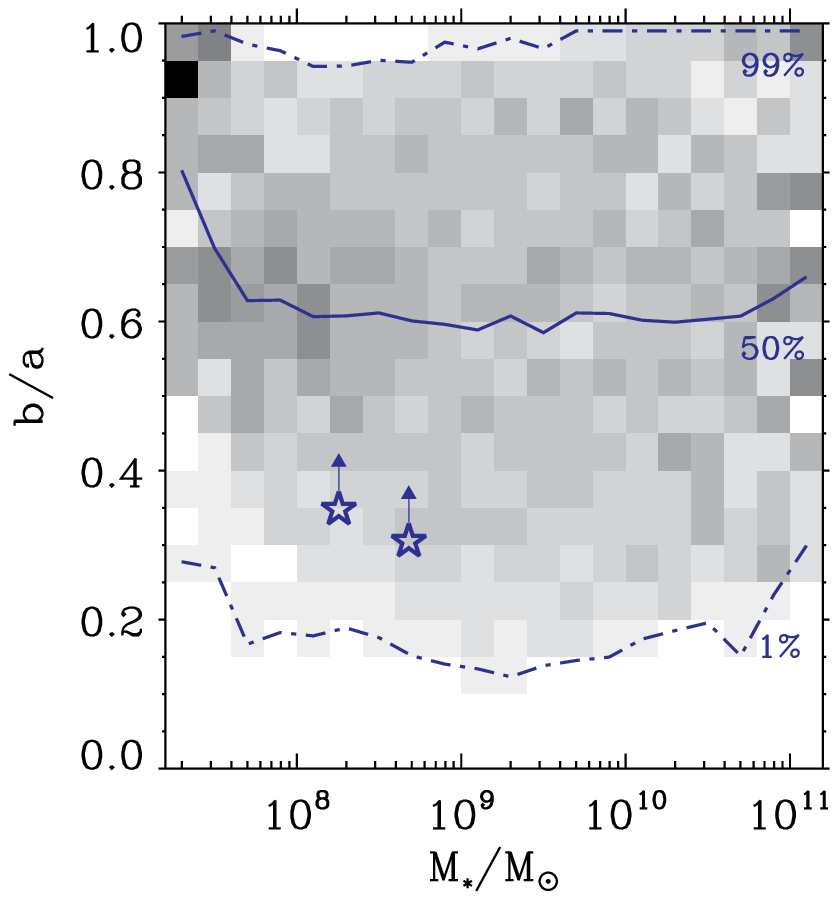}}
\resizebox{0.33\textwidth}{!}{\includegraphics{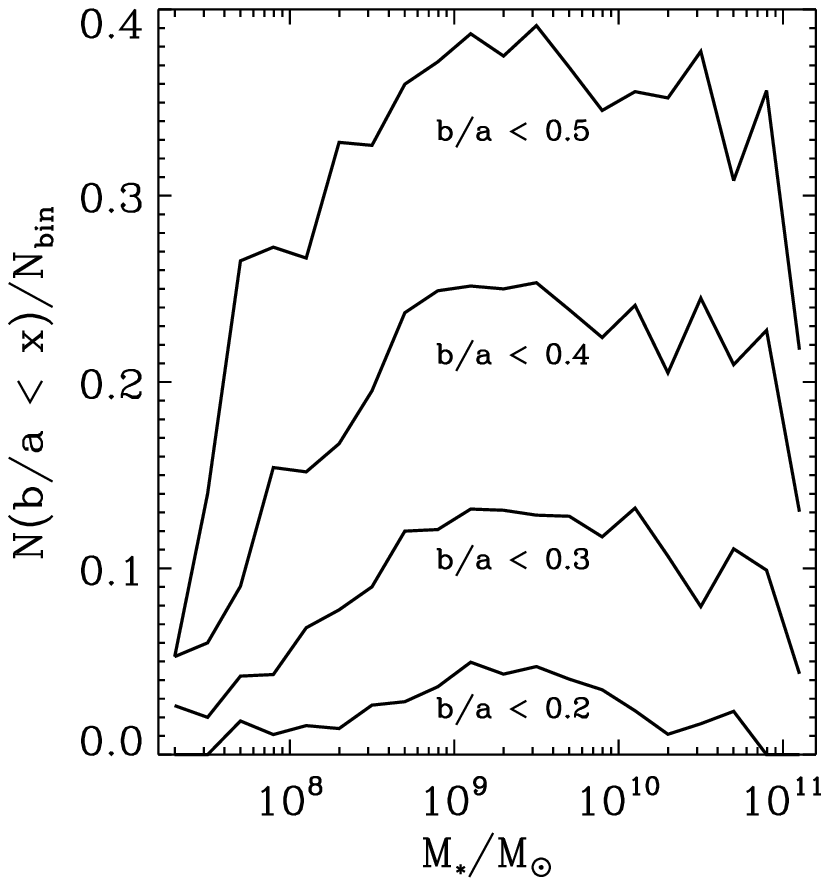}}
\resizebox{0.33\textwidth}{!}{\includegraphics{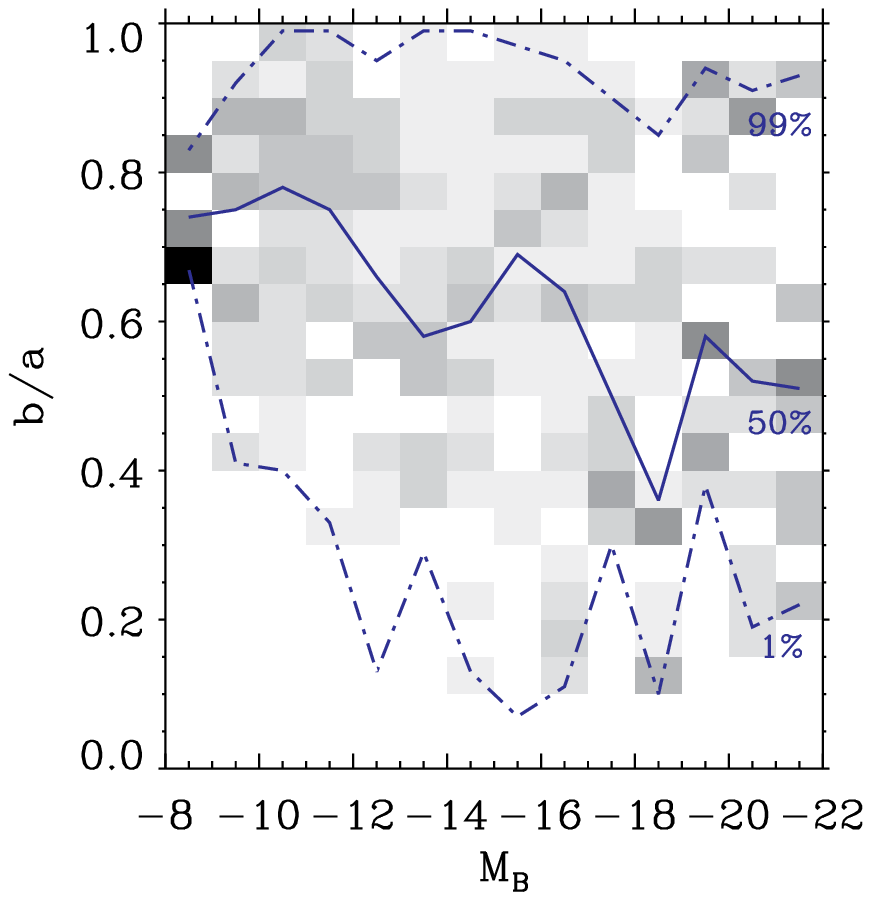}}
\caption{Probability distribution of $b/a$ in intervals of M$_{*}$ for the SDSS (left) and Local Volume (right) samples. The starred symbols indicate the edge-on thickness of the dwarf galaxies simulated by \cite{G10}. The central figure shows the fraction of galaxies in each mass interval having a lower $b/a$ than the specified value. The peak of the distribution at M$_{*} \approx 2\times10^{9}$ M$_{\odot}$ can be identified as the minimum mass of thin disc galaxies.}
\label{fig:1}
\end{figure*}

In Fig.\,\ref{fig:1} we present the probability distribution (grey scale) of $b/a$ in intervals of stellar mass (luminosity) for the SDSS (Local Volume) sample. The distribution is a strong function of M$_{*}$ ($M_{B}$), with M$_{*} \lesssim 2\times10^{9}$ M$_{\odot}$ galaxies being systematically thicker as they are fainter. The fraction of low-mass thin galaxies never exceeds a few per cent, and the overall $b/a$ distribution suggests that galaxies with stellar masses M$_{*} \sim 10^{7}$ M$_{\odot}$ are most probably spheroidal systems.

\subsection{Ruling out environmental effects}
\label{sec:2}

\begin{figure}
\resizebox{0.75\columnwidth}{!}{\includegraphics{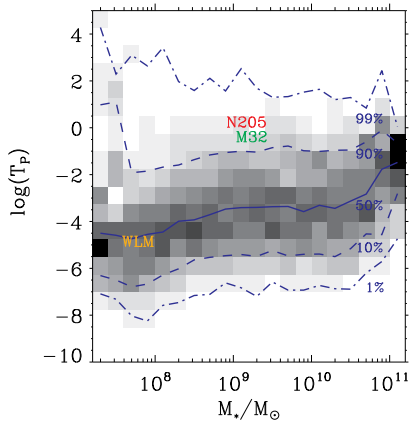}}
\caption{Probability distribution of the tidal parameter in bins of stellar mass for the SDSS sample. Note that low-mass galaxies are in general less tidally affected than their more massive counterparts. For reference we show the corresponding values for three local-volume dwarf galaxies experiencing different levels of interaction with M31.}
\label{fig:2}
\end{figure}

In order to test if environmental effects could be responsible of the observed trend, for each SDSS galaxy we have computed a tidal parameter that is proportional to the maximum ratio of external to internal forces that act on a galaxy \cite{SJ10}. We find that low-mass galaxies are in general less tidally affected than their more massive counterparts (see Fig.\,\ref{fig:2}) --consistent with them being mostly central and not satellite galaxies. We conclude that tidal effects are not responsible of the increasing thickness found in fainter galaxies, and suggest that the effect is probably related to the increasing importance of feedback mechanisms in low-mass haloes. 

\section{The effects of stellar feedback and their implications}
\label{sec:1}
Numerical and analytical estimates \cite{K07} suggest that the final morphology of a galaxy is the result of the complex interplay between the turbulent support provided by a gas temperature floor  --as might arise in the presence of a cosmic UV field and/or due to stellar feedback-- and the angular momentum support, characterised by the dimensionless spin parameter $\lambda$. For low-mass haloes, the pressure support radius becomes comparable to or larger than the rotational support radius, and thus galaxies are naturally formed thicker.

In order to check if stellar feedback effects can be effectively responsible of the previous trend, we have derived the intrinsic thickness of the dwarf galaxies recently modelled by \cite{G10}. These hydrodynamical simulations have first produced realistic dwarf galaxies thanks to their high resolution and a detailed treatment of baryonic processes.
We measured 25 mag\,arcsec$^{-2}$ isophotal axis ratios using $r$-band edge-on images of the dwarfs (starred symbols in Fig.\,\ref{fig:1}, left).
Interestingly, the least massive galaxy simulated by \cite{G10}, despite having the same $\lambda$ and a quieter merging history than the more massive one, is even thicker, nicely following the trend with mass we find. However,  their $q_{0}$ are higher than the minimum thickness we obtain for similar mass galaxies, and we speculate as to whether these flatter objects could reside in haloes with higher spin parameters.
We therefore suggest that star formation feedback effects --which can remove low angular momentum material and produce bulgeless dwarfs with shallow central dark matter profiles-- are also responsible of the increasing thickness of low mass galaxies. The shallower potential well certainly allows more turbulent (and therefore also vertical) motions, so stars naturally settle in a thicker disc. 

The fact that fainter galaxies are systematically thicker has further profound implications. First, the prevalence of turbulent motions likely inhibits the formation of bars and spirals in low-mass galaxies \cite{MA10}.  Second, a varying intrinsic thickness can introduce serious systematic errors in calculated inclinations if they are computed assuming discs have a universal, fixed $q_{0}$ --as is usually done when deprojecting H{\sc i} line profiles. Finally, thick dwarfs with shallow central dark matter profiles should be more susceptible to dynamical interactions than most simulations predict.

\end{document}